\documentclass[aps,twocolumn,showpacs]{revtex4}
\usepackage{graphics}
\usepackage{graphicx}

\begin{document}
\title{Anisotropic conductivity of doped graphene \\
due to short-range non-symmetric scattering}
\author{F.T. Vasko}
\email{ftvasko@yahoo.com}
\affiliation{Institute of Semiconductor Physics, NAS of Ukraine,
Pr. Nauki 41, Kiev, 03028, Ukraine }
\date{\today}

\begin{abstract}
The conductivity of doped graphene is considered taking into account scattering
by short-range nonsymmetric defects, when the longitudinal and transverse
components of conductivity tensor appear to be different. The calculations of the
anisotropic conductivity tensor are based on the quasiclassical kinetic equation
for the case of monopolar transport at low temperatures. The effective longitudinal
conductivity and the transverse voltage, which are controlled by orientation of
sample and by gate voltage (i.e. doping level), are presented.
\end{abstract}


\maketitle
The conductivity tensor $\sigma_{\alpha\beta}$ is determined both by the symmetry
properties of carriers and by the character of scattering processes. For a medium with
cubic (quadratic for the 2D case) symmetry, the conductivity appears to be a scalar,
i. e. $\sigma_{\alpha\beta}\propto\delta_{\alpha\beta}$. \cite{1} Anisotropic
conductivity of hot electrons in low-symmetric bulk materials (Ge and Si) were
studied over 50 years ago. \cite{2} For the hexagonal symmetry case, which is correspondent to an ideal graphene sheet, the longitudinal (along $X$-axis, see
Fig. 1) and transverse conductivities of {\it hot carriers} appear to be different.
This anisotropic conductivity appears due to the trigonal warping of energy spectrum \cite{3} even for the isotropic scattering case. Such an anisotropy is weak
[$\propto (a\bar{p}/\hbar )^2$, where $a$ is the lattice constant and $\bar{p}$ is
the characteristic momentum] and it can be essential for the high-energy carriers,
at energies $\geq 1.5$ eV. {\it In the linear regime}, the anisotropy of conductivity
is possible due to scattering by short-range nonsymmetric defects, see Fig. 1b
where a substitutional impurity is shown. Although the nonsymmetric defects were
discussed in Refs. 4-7, the anisotropy of conductivity was not theoretically analyzed
or mentioned. To date, no experimental data are reported concerning any mechanism
of anisotropy listed; so that, a study of this phenomena is timely now.

\addvspace{-1 cm}
\begin{figure}[ht]
\begin{center}
\includegraphics{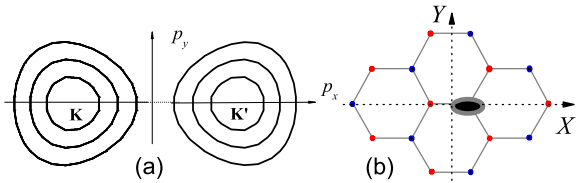}
\end{center}
\addvspace{-1 cm}
\caption{(Color online) (a) Equal enegry contours around the $K$ and $K'$ points
with step 0.5 eV; a visible anisotropy takes place at energies $\geq 1.5$ eV.
(b) Graphene sheet with a short-range nonsymmetric defect marked by ellipse.}
\end{figure}

In this Letter, we consider the low-temperature conductivity taking into account
both the long-range isotropic disorder (see \cite{8,9,10}) and the short-range
nonsymmetric defects (see \cite{4,5,6,7}). The longitudinal and transverse components
of conductivity tensor are calculated based on the linearized Boltzmann equation
with the transition probabilities written in the Born approximation. The effective
conductivity and the transverse voltage of a graphene strip are presented for the case
of weak anisotropy and their dependencies on carrier's concentration (gate voltage)
and on orientation of strip are discussed.

We consider the momentum relaxation in graphene which is caused by the interaction
of carriers with a weak potential $\sum_j\hat U({\bf x}-{\bf x}_j)$, where ${\bf x}_j$
is the coordinate of $j$th impurity ($j=1,\ldots N_{im}$ and $N_{im}$ is the number
of impurities over the normalization area $L^2$). The potential of a single impurity
placed at ${\bf x}=0$ is given by $\hat U({\bf x})=v_{\bf x}+\hat u\delta ({\bf x})$;
here we separated the long-range scalar contribution, $v_{\bf x}$, and the short-range
addend determined by the $4\times 4$ matrix $\hat u$. For a nonmagnetic defect,
this matrix should be Hermitian and time-reversal symmetric and one can write
$\hat u$ using 9 independent real parameters $u_{sl}$:
\begin{equation}
\hat u=\sum\limits_{s,l=x,y,z}u_{sl}\hat\Sigma_s\hat\Lambda_l ,
\end{equation}
where $4\times 4$ matrices $\Sigma_s$ and $\Lambda_l$ are introduced in Refs. 4-6.
In the framework of the Born approximation, the transition probability between
the $c$-band states $|\eta '{\bf p}'\rangle$ and $|\eta{\bf p}\rangle$ is given by
\begin{equation}
W_{\eta '{\bf p}'\eta {\bf p}}=\frac{2\pi}{\hbar}\frac{n_{im}}{L^2}\left|\left
\langle\eta '{\bf p}'\left|{\hat U}\right|\eta {\bf p}\right\rangle \right|^2
\delta\left(\varepsilon_{p'}-\varepsilon_p\right) ,
\end{equation}
where $\bf p$ is 2D momentum, $n_{im}=N_{im}/L^2$ is the impurity concentration,
$\eta =K,K'$ is correspondent to $K$- or $K'$-valleys
and the linear dispersion law $\varepsilon_p=v_Wp$ is not dependent on $\eta$
($v_W=10^8$ cm/s is the characteristic velocity). \cite{3} Below we restrict
ourselves by the case of weak short-range corrections to the long-range scattering
contribution.
As a result, one obtains the intravalley matrix element in (2) as
\begin{eqnarray}
\left|\left\langle K{\bf p}'\left|{\hat U}\right| K{\bf p}\right\rangle\right|^2
\simeq v_q^2\frac{1-\cos\Delta\varphi}{2} \\
+\frac{u_{xz}^2}{2}\left[ 1+\cos\left(\varphi +\varphi '\right)\right] +
\frac{u_{yz}^2}{2}\left[ 1-\cos\left(\varphi +\varphi '\right)\right] , \nonumber
\end{eqnarray}
where the main contribution is written through the Fourier component of $v_{\bf x}$,
which is dependent on the momentum transfer $\hbar{\bf q}={\bf p}-{\bf p}'$, and
through the factor $(1-\cos\Delta\varphi )$ with $\Delta\varphi =\widehat{{\bf p}',
{\bf p}}$ described the suppression of backscattering processes. The
$\Delta\varphi$-dependent contributions, which are proportional to $u_{sl}^2$, were
omitted in Eq. (3). In addition, a weak $\Delta\varphi$-dependent intervalley matrix
element, which is proportional to $u_{zl}^2$ or to $u_{sl}^2$ with $(s,l)\to (x,y)$
should also be neglected. The remaining corrections in Eq. (3), which are dependent
on $\varphi +\varphi '$ ($\varphi$ and $\varphi '$ are the polar angles of $\bf p$
and ${\bf p}'$), are responsible for the anisotropic conductivity under consideration
if $u_{xz}^2\neq u_{yz}^2$, i.e. scattering processes along $X$- and $Y$-directions
are different.

Next, we solve the Boltzmann kinetic equation linearized with respect to the
steady-state electric field $\bf E$. Taking into account that the transition
probability given by Eqs. (2), (3) does not dependent on $\eta$ and that
$W_{\bf pp'}=W_{\bf p'p}$, we write the kinetic equation for the anisotropic
part of distribution $\Delta f_{\bf p}$, which is $\propto {\bf E}$, in the
following form:
\begin{equation}
e\left({\bf E}\cdot{\bf v}_{\bf p}\right)\frac{df_\varepsilon}{d\varepsilon} =
\sum\limits_{{\bf p}'}W_{{\bf p}'{\bf p}}\left(\Delta f_{{\bf p}'}-\Delta f_{\bf p}
\right) .
\end{equation}
Here ${\bf v}_{\bf p}=\nabla_{\bf p}\varepsilon_p$ is the velocity, $f_\varepsilon$
is the equilibrium distribution and $df_\varepsilon /d\varepsilon\simeq -\delta
(\varepsilon_F -\varepsilon_p )$ for the degenerate carriers with the Fermi energy $\varepsilon_F$. We solve the integral equation (4) using the variational approach
\cite{11} with the trial distribution function $\Delta f_{\bf p}=e\left({\bf C}\cdot
{\bf v}_{\bf p}\right)\delta (\varepsilon_F -\varepsilon_p)$. An unknown vector
$\bf C$ is determined from the extremum conditions for the quadratic form:
\begin{equation}
{\cal K}({\bf C})=\left({\bf C}\cdot\hat A\cdot{\bf E}\right)+\left({\bf C}\cdot\hat L
\cdot{\bf C}\right) ,
\end{equation}
where the field-dependent contribution is determined by the tensor:
\begin{equation}
A_{\alpha\beta}=\frac{2\upsilon_W^2}{L^2}\sum\limits_{\bf p} {\delta (\varepsilon_F
-\varepsilon_p )\frac{{p_\alpha  p_\beta  }}{{p^2 }} = \delta _{\alpha \beta } }
\frac{\upsilon_W^2\rho_F}{4}
\end{equation}
written through the density of states at the Fermi energy, $\rho_F$.
The collision-induced contribution in Eq. (5) is determined by the matrix:
\begin{eqnarray}
L_{\alpha\beta}=\frac{2\pi n_{im}\upsilon_W^2}{{\hbar L^4 }}\sum\limits_{{\bf pp}'}
\left|\left\langle K{\bf p}'\left|{\hat U}\right| K{\bf p}\right\rangle \right|^2
\delta (\varepsilon_F-\varepsilon_p )  \nonumber \\
\times\delta (\varepsilon_F-\varepsilon_{p'})\frac{p_\alpha (p'_\beta -p_\beta )}
{p^2 }
\end{eqnarray}
and after substitution of the matrix elements (3) one obtains $L_{\alpha\beta}=0$ if $\alpha\neq\beta$.
The diagonal components of Eq. (7) take form:
\begin{equation}
L_{\alpha\alpha}\simeq\frac{\pi n_{im}\upsilon_W^2\rho_F^2v_0^2}{8\hbar}\left[\Psi
\left(\frac{p_F l_c}{\hbar}\right)\pm\frac{u_{xz}^2 -u_{yz}^2}
{8v_0^2} \right] ,
\end{equation}
where $+$ (or $-$) is correspondent to $L_{xx}$ (or $L_{yy}$). The factor
$v_0^2\Psi (p_F l_c/\hbar )$ appears here due to the long-range part of potential
($v_0\equiv v_{q=0}$) with the correlation lenth $l_c$. The function $\Psi
(p_F l_c/\hbar )$ was introduced in \cite{9} for the case of the finite-range
potential with Gaussian correlations.

The current density is given by ${\bf I}=(4e/L^2) \sum_{\bf p}{\bf v}_{\bf p}\Delta
f_{\bf p}$, where the factor 4 is due to the spin and valley degeneracy. Since
$\Delta f_{\bf p}\propto{\bf C}\propto{\bf E}$, the conductivity tensor $\hat\sigma$
is introduced as ${\bf I}=\hat\sigma {\bf E}$. Substituting $\Delta f_{\bf p}$ into
$\bf I$ and performing integration over $\bf p$, one obtains ${\bf I}=(ev_W )^2{\bf C}
\rho_F /2$, where $\bf C$ is determined from the extremum conditions which give the
linear equations: $\partial {\cal K}({\bf C})/\partial C_\alpha =0$ for $\alpha =x,y$.
As a result, one obtains the diagonal tensor $\sigma_{\alpha\alpha}=\overline{\sigma}\pm\delta\sigma /2$, where $+$ (or $-$) is correspondent to $\sigma_{xx}$ (or
$\sigma_{yy}$). Here the isotropic part of conductivity $\overline{\sigma}$ and
the anisotropic correction $\delta\sigma$ are given by
\begin{eqnarray}
\overline{\sigma}\simeq\frac{e^2}{\hbar}\frac{(\upsilon_W\hbar )^2}{2\pi n_{im}v_0^2
\Psi (p_Fl_c /\hbar)}, \nonumber \\
\delta\sigma\simeq -\frac{\overline{\sigma}}{\Psi (p_Fl_c/\hbar )}
\frac{u_{xz}^2-u_{yz}^2}{8v_0^2} ,
\end{eqnarray}
so that the anisotropy of conductivity is determined by the dimensionless factor
$(u_{xz}^2-u_{yz}^2) /(8v_0^2)$.

\begin{figure}[ht]
\begin{center}
\includegraphics{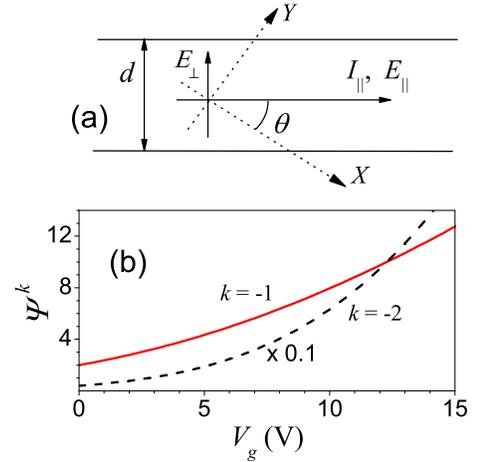}
\end{center}\addvspace{-1 cm}
\caption{(Color online) (a) Graphene strip of width $d$ with angle $\theta$
to $X$-direction (dotted arrows are correspondent to the $X0Y$ coordinate system).
Longitudinal and transverse fields, $E_{\bot}$ and $E_{\|}$, as well as
longitudinal current $I_{\|}$ are shown. (b) Functions $\Psi^{-1}$ and
$\Psi^{-2}$ which describe gate voltage dependencies of transverse voltage,
$E_\bot /E_{\|}$, and of anisotropic contribution to conductivity, $(\sigma_{eff}
-\overline{\sigma})$, given by Eq. (10).}
\end{figure}

Using the conductivity tensor determined by Eqs. (9), we consider the conductivity
of strip, see Fig. 2a where $d$ is the width of strip and $\theta$ is the angle
between its orientation and $X$-axis. Since the absence of the transverse current,
$I_{\bot}=0$, the linear relation ${\bf I}=\hat\sigma{\bf E}$ permits one to
calculate the effective conductivity, which is introduced according to $I_{\|}=
\sigma_{eff}E_{\|}$, and the induced transverse voltage, $E_{\bot}d$. For the weak
anisotropy case, we obtain the simple angle dependencies
\begin{equation}
\frac{E_\bot}{E_{\|}}\simeq\frac{\delta\sigma}{\overline{\sigma}}\sin 2\theta , ~~~~
\sigma_{eff}-\overline{\sigma}\simeq\frac{\delta\sigma}{2}\cos 2\theta
\end{equation}
and maximal values of the transverse voltage $|E_\bot /E_{\|}|$ and of the anisotropic contributions to $\sigma_{eff}$ appear at $\theta =\pi /4$, $3\pi /4,\ldots$ and
at $\theta =0$, $\pi /2,\ldots$, respectively.
The concentration (or gate voltage, $V_g$) dependencies are determined through the
function $\Psi (p_Fl_c /\hbar)$ introduced in Eqs. (8), (9). According to Eq. (10),
one obtains $|E_\bot /E_{\|}|\propto\Psi^{-1}$ and $\delta\sigma\propto\Psi^{-2}$;
we plot these functions in Fig. 2b for the graphene strip placed over the SiO$_2$
substrate of 300 nm width. We also use $l_c\simeq$10 nm in agreement with experimental data, see Ref. 9. Since the function $\Psi$ decreases with $V_g$, both the transverse
voltage $|E_\bot /E_{\|}|$ and the anisotropy of conductivity $(\sigma_{eff}-
\overline{\sigma})$ increase with $V_g$; notice, that the anisotropy of conductivity
increases more faster (about 10 times at $V_g\simeq$10 V) because $\delta\sigma
\propto\Psi^{-2}$. The transverse field and the anisotropy of effective
conductivity given by Eq. (10) are proportional to the unknown parameter $(u_{xz}^2-
u_{yz}^2)/(8v_0^2)$ and a value of the anisotropy effect is not estimated here.
However, these results permit one to verify a mechanism of anisotropy using the gate voltage and angle dependencies obtained. In addition, a ratio $(u_{xz}^2-u_{yz}^2)/(8v_0^2)$ can be measured. Supposing $(u_{xz}^2-u_{yz}^2)/(8v_0^2)\sim 10^{-2}$,
one obtains that $|E_\bot /E_{\|}|$ and $|\delta\sigma /\overline{\sigma}|$ are
about 10 \%.

Further, we discuss the assumptions used. The consideration performed is based on
the simplified description of scattering with separated long- and short-range
parts of potential. The parameters of long-range scattering are determined from
the phenomenological consideration \cite{9} while the short-range contribution
is expressed through unknown matrix $\hat u$, without verification of a
microscopical nature of defects. The variational solution of Eq. (4) gives a good
estimation of the anisotropy under consideration. The results (10) do not dependent
on a width of strip if $d$ exceeds the mean free path due to the long-range scattering,
so that an edge scattering is not essential under the condition $d\gg$0.1 $\mu$m
(here we used the results for non-ideal nanoribbons \cite{12}). We also
restrict ourselves by the degenerate carriers case because the anisotropy effects
increase with concentration (or gate voltage). The rest of assumptions
(quasiclassical kinetic equation, weak acoustic phonon scattering, and model
description of long-range scattering \cite{9}) are rather standard.

In conclusion, we have demonstrated that the nonsymmeric short-range scattering
gives rise to the transverse voltage and to the orientation-dependent
effective conductivity of graphene strip. These dependencies permit one to
measure asymmetry of scattering which is expressed through the characteristics of
short-range defect. We believe that these results will stimulate further measurements
of anisotropy and microscopical calculations of nonsymmetric defects. Beside this,
a nonequivalent heating of carriers in different valleys (and an intervalley
redistribution of carriers) by a strong electric field takes place under an
essential anisotropy of scattering (in analogy with the bulk case \cite{13}).

I am grateful to E. I. Karp for a useful comment.

\end{document}